# Evidence for polarons in iron pnictides of the *Ln*-1111 and *Æ* -122 families

Manuel Núñez-Regueiro

Institut Néel, Centre National de Recherches Scientifiques & Université Joseph Fourier, BP166 cedex 9, 38042 Grenoble, France

Examination of the electrical resistivities of iron pnictides shows that they can be accounted by conduction by polarons. Their activation energies show a linear behaviour with the critical temperatures of the spin density waves (SDW), $T^*$, as both vary with pressure. The slope matches the ratio SDW gap to $T^*$, while the intercept can be related to the transition temperature of the lattice distortion, $T_0$. An adapted Landau free energy predicts the observed order of the transitions, according to which is higher, $T^*$ or $T_0$. Simple arguments favour combined Jahn-Teller antiferromagnetic bipolarons.

email: nunez@grenoble.cnrs.fr

One of the main issues in superconductivity is the origin of the force binding electrons in pairs that condense into the quantum macroscopic state. High temperature superconductors were discovered following the belief that Jahn-Teller polarons [1] could render strong binding forces, though the precise origin of the pairing interaction in cuprates still remains controversial. A new opportunity has now appeared with the report of superconducting transition temperatures up to ~56K in compounds with iron in tetrahedral coordination[2, 3, 4, 5]. As in cuprates, the ground state of the mother compounds exhibits antiferromagnetic order[6], whose residual interactions on doping are presently held responsible for the pairing. However, in contrast with cuprates it is an itinerant antiferromagnetic state of the spin density wave type (SDW)[7]. However, the lattice distortion[8] that is also always present in the ground state nor its relation to the SDW are presently understood. Besides, the reason why the phase transtion towards the antiferromagnetic state is always second order, while the lattice distortion can be second order for *LnFeAsO* (*Ln-1111*), where *Ln* is a lanthanide, or first order for *ÆFe$_2$As$_2$* (*Æ-122*), where *Æ* is an alkaline-earth, remains also a puzzle. More basically, the "bad metal" electrical resistance of these materials at high temperatures, lacks a coherent and straightforward interpretation.

Here it is shown that the electronic transport properties of the mother compounds can be naturally explained assuming the existence of polarons at high temperatures, whose cooperative ordering can bring about the lattice distortion. The analysis of the correlation obtained between the activation energy of the polarons and the transition temperature towards the SDW, $T^*$, leads to the existence of antiferromagnetic bipolarons. The strong coupling between the lattice and antiferromagnetic transitions is analyzed and the order of the phase transitions deduced, in agreement with experiment.

Band structure calculations for the undoped *LnFeAsO* (*Ln*-1111) oxypnictides yield [9] two dimensional cylindrical hole and electron Fermi surfaces that are prone to nesting, rendering an itinerant antiferromagnetic ground state in the form of a SDW7, as is found experimentally[6] at $T^*$. The archetype for this type of phase transition is chromium metal[10], which develops a SDW at $T^*=312K$. However, the dependence of the electrical resistance of chromium, metallic above and below $T^*$, though common to most of SDW compounds (e.g. Cobaltites[11]), is strikingly different to the dependence observed in pnictides. As a tetragonal to orthorhombic lattice distortion (TOLD) is also observed[8] at a transition temperature $T_0 > T^*$, it is basic to take into account possible electron-lattice coupling effects. As the crystal field effect of the arsenic tetrahedron on the five degenerate 3d Fe levels can explain several of the *Ln*-1111 properties[12], it is natural to refer to materials with similar configurations, such as manganites[13]. In fact, the resistivity curve on Fig 1(a) bears many resemblances with the behaviour observed in those materials[13], where the transport at high temperatures is attributed to small Jahn-Teller polarons with a $\rho \sim T\exp(E_a/k_B T)$ temperature dependence, $\rho$ being the resistivity, $E_a$ an activation energy, and $k_B$ the Boltzmann constant. Applying the assumption of polaronic transport on pnictides, the subsequent Arrhenius plot of Fig 1(b) allows determining $E_a$ for the *Sm*-1111 oxypnictide. Although this sample is polycrystalline an identical behaviour is observed for the monocrystalline *Sr*-122 pnictide, Fig. 1(c) and Fig 1(d), thus confirming that it is an intrinsic feature. $E_a$ can thus be obtained from the Arrhenius of the published data on *Ln*-1111 and Æ-122 pnictides electrical resistivity data, while $T^*$ can be determined from the peak of the derivative of the resistivity with respect to temperature, $d\rho/dT$. Both parameters are plotted on Fig 2 a and Fig 2 b as a function of the basal lattice parameter resulting in a linear correlation for all the rare earths (with the exception of *La*, whose doped samples also have an anomalously low superconducting transition temperature) on the *Ln*-1111 plot and for Ca and Sr on the Æ-122 plot. Remarkably, the $E_a$ has a magnitude comparable to the Fe 3d level separation in the distorted tetrahedral field[12]. It is tempting to associate these polarons to a Jahn-Teller (JT) effect, since, as shown on Fig 3d, there is splitting of a doublet when the TOLD takes place. JT polarons may exist above $T_0$, when they would condensate into a permanent TOLD, through, presumably, a cooperative JT transition[14]. Alternatively, polarons have been recently[15] predicted in pnictides, although based on the highly polarizable $As\ 4p$ states, and not on crystal field JT effects.

The optimal way to determine the relationship between $E_a$ and $T^*$ is from pressure measurements on the same sample[16], as in such a way the differences between different samples (defects, impurities, etc.) are avoided. Up to date reported pressure measurements are found on $Sm$ and $La$ -1111 and $Ca$, $Sr$ and $Ba$-122. The activation energy and the transition temperature for each pressure are now extracted from the published data, and plotted one against the other on Fig 2c and Fig 2d, each point being one different pressure. The dependences are extremely linear, i.e. $E_a = \varepsilon + \alpha T^*$, with the same slope $\alpha$ for each family materials but different intercept energy $\varepsilon$ for each compound. Referring once more to manganites, it is useful to propose a band picture for the strongly coupled electron-phonon system (similar to the one on Fig. 1 of Ref. 17) including narrow polaronic bands (Fig 3 c). The point here is that, contrary to manganites, there is no real gap in pnictides. However, precursor fluctuations of the SDW, rendering a pseudogap above $T^*$, are currently expected (due to a nematic phase[18] or dimensionality fluctuations[19]). Also, a pseudogap has been used to explain the anomalous magnetic susceptibility[20] of undoped pnictides at $T > T^*$. Thus, here a pseudogap $\Delta^*$ replaces the band gap of manganites, with the smaller polaronic gap between electron and hole polaron bands. This latter gap $\varepsilon_0$ can be obtained through thermoelectric power measurements[13,17]. The thermopower of small polaronic systems is similar to that of band semiconductors, governed by thermal activation of carriers across a small barrier and thus a function of the inverse temperature: $S = \dfrac{k_B}{e}\left(\dfrac{E_S}{k_B T} + b\right)$ where $S$ is the thermopower, $E_S$ is an activation energy ($=\varepsilon_0$), $e$ the electronic charge and $b$ other, negligible terms. The value of the activation energy $E_S$ for several *Ln*-1111 samples is thus determined from the measured thermopowers through the plot on Fig 3 a. It can now be compared to the intercept $\varepsilon$ (Fig 3 b). To a good approximation $\varepsilon \approx E_S = \varepsilon_0$, i.e. as a first estimate the intercept $\varepsilon$ corresponds to the polaronic gap $\varepsilon_0$.

Analyzing now the activation energy of the electrical resisitivty, for small polarons it is, $E_a = \varepsilon_0 + W_H - J$, where $W_H$ is one half[13,21] of the polaron formation energy $E_P$, and $J$ the transfer integral. As from the diagram on Fig 4c $\Delta^* = \varepsilon_0 + E_P$, then $E_a = \frac{\varepsilon_0 + \Delta^*}{2} - J$. The relation between the gap and the transition temperature in a mean field approximation is $\Delta = 1.75 k_B T^*$, and more generally $\Delta^* = \delta k_B T^*$. Thus, $E_a = \frac{\varepsilon_0 + \delta k_B T^*}{2} - J$ (1). In this way the linear dependence of $E_a$ on $T^*$ can be explained, with the slope $\alpha$ related to the ratio gap to critical temperature, $\delta$. However, from this expression the intercept $\varepsilon$ should be $\sim E_S/2 = \varepsilon_0/2$, when it is clear from Fig 4b, that $\varepsilon \approx E_S = \varepsilon_0$. Although $\varepsilon_0$ can be expected to change with pressure, this variation may be included into the actual error of determination, i.e. 20%. The ratio $\delta$ of the gap to the critical temperature is obtained from the reported optical absorption gap measurements for *Ln*-1111 [22] and *Æ*-122 [23] materials, yielding $\delta_{1111} \approx 1.04 \pm 0.4$ and $\delta_{122} \approx 1.75 \pm 0.05$. These values are almost identical to the slopes $\alpha_{1111} = 1.16 \pm 0.1$ and $\alpha_{122} = 1.60 \pm 0.06$, while according to formula (1) they should be twice their value. Thus, the precedent analysis agrees qualitatively and quantitatively with experiments if the factor ½ is eliminated, i.e. if electronic transport is performed by two polarons simultaneously, in a first approximation bipolarons. This ½ factor that points towards bipolarons appears in the formulas through the relation that states that the hopping of polarons costs half their energy of formation, relation that has been recurrently time-tested[13,21].

It is interesting to note that while the intercept $\varepsilon$ for the *Ln-1111* compounds is positive, that for *Ba*-122 is also positive but very small, and is negative for *Sr* and *Ca*, meaning that polarons are deep inside the Fermi sea for the later two. It is reasonable to assume that, while long lifetime polarons (*Ln*-1111 and eventually *Ba*-122) can trigger a TOLD before the appearance of the SDW, i.e. $T_0 \geq T^*$, those with a very short lifetime through strong hybridization (*Sr*, *Ca*) will have very low or nil condensation temperatures and in this case $T^* \gg T_0$. In other words, the intercept $\varepsilon$ should be related to the part of the activation energy that is due to the JT deformation, positive for the *Ln-1111* compounds and thus increasing their stability, while for the *Æ*-122 it is almost zero or negative and decreasing their stability.

A Landau free energy analysis is useful here to study the scenario of the interrelation between the transitions, in particular the order of the phase transformations. According to specific heat measurements, the phase transitions are of second order for the *Ln*-1111 ($\varepsilon>0$) and *Ba-122* ($\varepsilon>\sim 0$)[24] (although for this last compound this seems to be sample dependent[25], as would be expected due to its very small $\varepsilon$), and of first order for *Sr-122* [26] and *Ca-122* [27]. The Landau free energy for the system reads

$$F = a(T - {}^uT^*)\eta_{SDW}^2 + a'(T - {}^uT_0)\eta_{LD}^2 + b\eta_{LD}\eta_{SDW}^2 - c\eta_{LD}^2\eta_{SDW}^2 + d\eta_{SDW}^4 + d'\eta_{LD}^4 \quad (2)$$

where $a, a', b, c, d, d'$ are parameters, ${}^uT^*$ and ${}^uT_0$ the uncoupled transition temperatures and $\eta_{LD}$ and $\eta_{SDW}$ are the order parameters for the TOLD and the SDW, respectively. The first, second, fifth and sixth terms are standard. The third linear-quadratic one is a particular coupling allowed by symmetry, as the wavevector for the TOLD is twice the one of the SDW (this term is common in spin-Peierls systems) and the fourth one explicits a strong favorable coupling that is assumed between the two transitions based on the same bands. The coupling translates the fact that due to the SDW transition, the bands degenerate at M and Γ split[9], as in a band JT transition[28]. Solving first for $T_0 \geq T^*$, putting as for the *Ln-1111* case ${}^uT_0 = 155K$ and ${}^uT^* = 90K$ (the expected transition temperature from the ratio $\delta_{1111}/1.75$ if mean field holds), both transitions are second order and the actual transitions temperatures $T_0 = 155K$ and $T^* = 145K$, in agreement with measurements (Fig 4 a), i.e. due to the coupling with the TOLD the SDW transition increases in temperature. While if the SDW appears before the TOLD, i.e. $T^* > T_0$, putting tentatively "$T_0 \sim 0K$ and "$T^* = 200K$, as expected for *Sr-122*, the SDW transition is second order but the TOLD transition is now first order, as is precisely found experimentally[29], with $T^* = 200K$ and (Fig. 4 b) $T_0 = 185K$, very near to the observed values (the same $a, a', b, c, d, d'$ have been used for both calculations). The order of the phase transitions is thus the immediate consequence of the expected by symmetry linear-quadratic coupling of the SDW-TOLD states.

A structure for the bipolaron would be nearest neighbours polarons coupled antiferromagnetically (Fig 4c), in a minimalist version of the TOLD-SDW. For *Sr*-122 and *Ca*-122 compounds, the energy gain from the antiferromagnetic coupling will stabilize the otherwise unstable polarons, while the distortion is the only way to create a magnetic moment ($\sim 0.35\mu_B$, Ref.12) on the Fe atom. The supposition of antiferromagnetic bipolarons would also justify the existence of a pseudogap at $T > T^*$. It must be noted here that these are not standard bipolarons, as the latter are singlet coupling of two polarons that have no localized spin. Transport at $T > T^*$ would be performed through antiferromagnetic bipolarons (BP) and free carriers (FC). FC would have a short constant mean free path due to scattering against the disordered BP and a comparatively small $T$ dependence, while the certainly pinned BP can be excited through $E_a$ to FC and re-pinned elsewhere, i.e. $\rho_{BP} \sim T\exp(E_a/k_BT)$. Below $T^*$, BP order into the coupled TOLD-SDW, while the ungapped FC would have a mean free path increasing with the TOLD-SDW gap temperature dependence.

In conclusion, it is shown that polaronic formation can explain the "bad metal" electrical resistivity of the mother compounds of the superconducting iron pnictides. From the analysis of the variation with pressure of the activation energy of the polarons and the SDW transition temperature, it is concluded that they are coupled in pairs of antiferromagnetically coupled Jahn-Teller bipolarons. Bipolarons in the mother compounds can be crucial in the understanding of the pairing mechanism of pnictides[30,15], if their existence were attested also in the doped materials. Besides, the fact that both imbricated transitions yield bipolarons strongly suggests that pairing should contain elements of both interactions. Finally, it is also demonstrated that the order of the SDW and TOLD transitions is the consequence of their particular linear-quadratic coupling within a Landau free energy analysis.

The author acknowledges Gastón Garbarino, Marie-Bernardette Lepetit, José Emilio Lorenzo, Pascal Quemerais and Rubén Weht for help during analysis, and Blas Alascio, Michel Avignon, Claudine Lacroix, Pierre Monceau and Julius Ranninger for fruitful discussions.

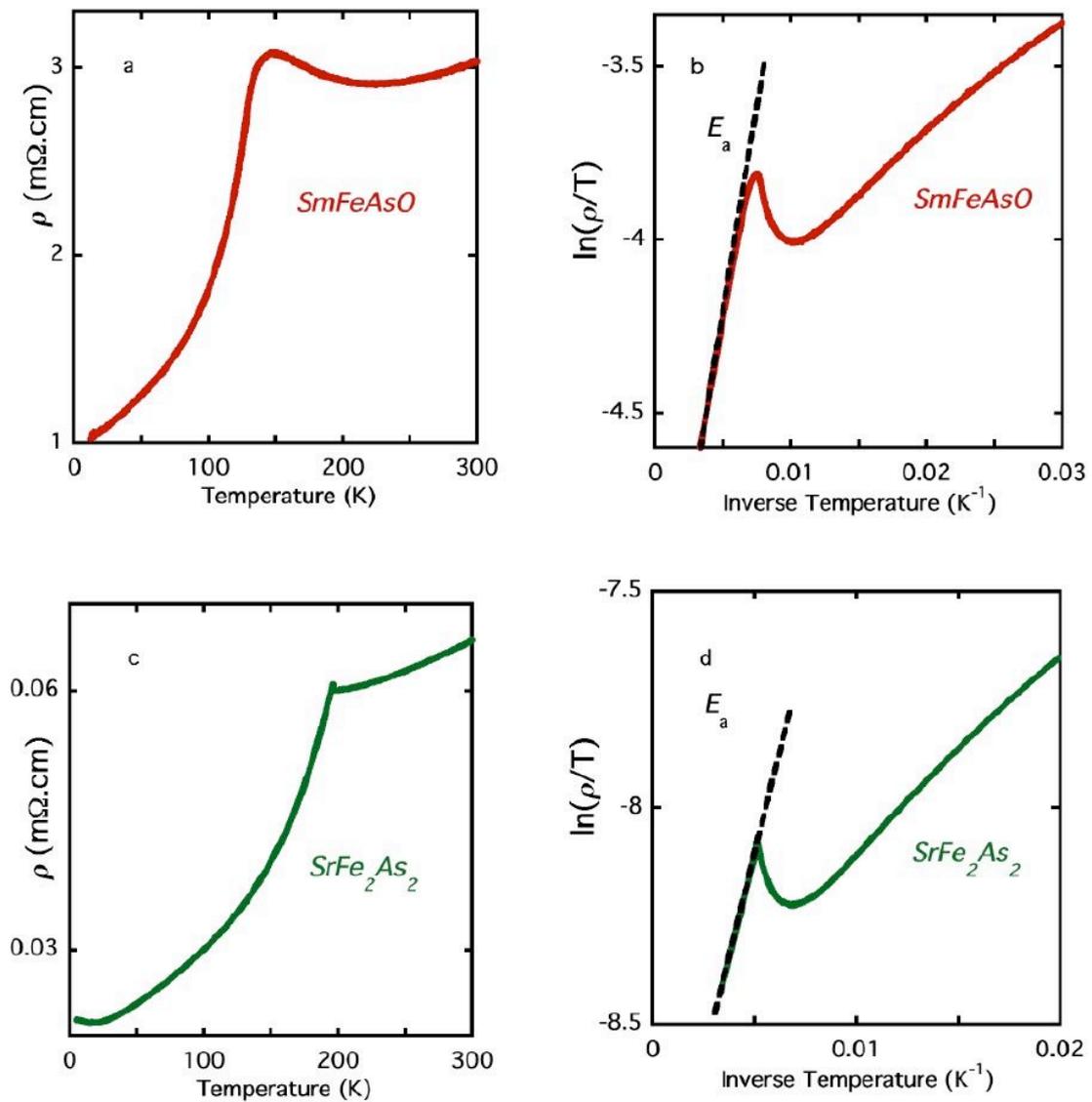

Figure 1. (a) Electrical resistance of a *SmFeAsO* sample[38]. (b) Data of (a) plotted in an Arrhenius plot (logarithm of resistivity/temperature versus inverse temperature) in order to determine the activation energy $E_a$ due to small polaron transport (c) Electrical resistivity of a *SrFe$_2$As$_2$* sample[31] (d) Same as (c) for the *SrFe$_2$As$_2$* sample.

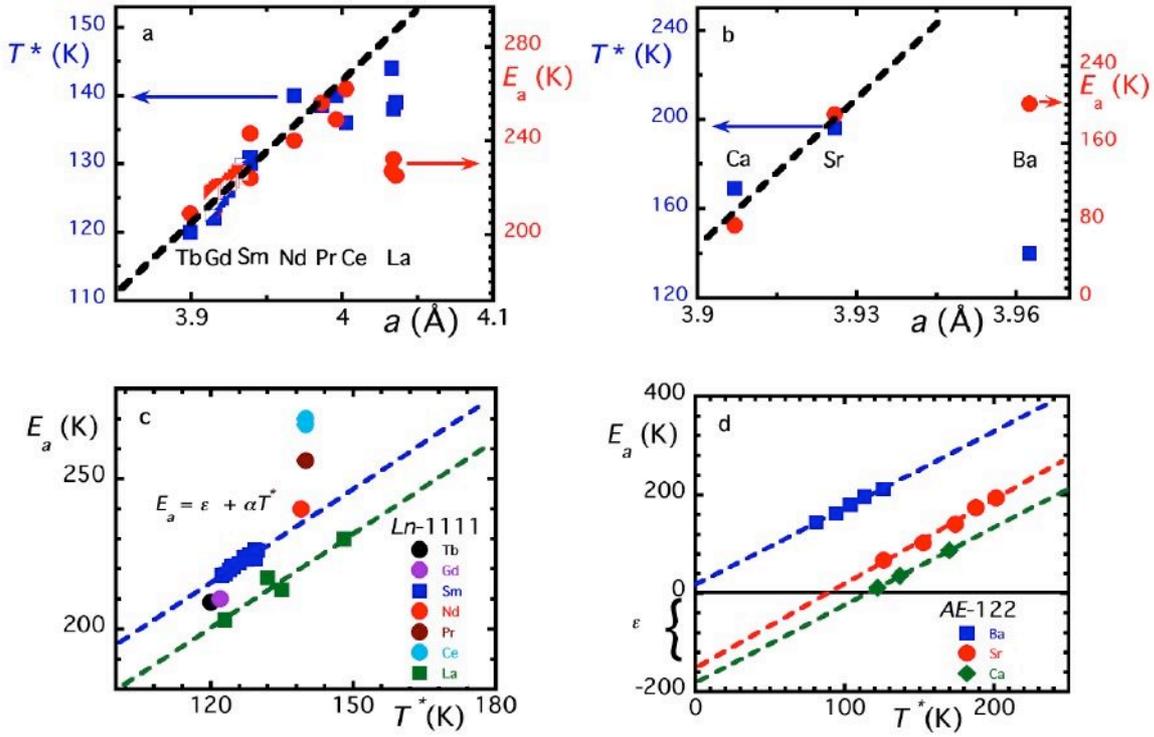

Figure 2 (a) Transition temperature $T^*$ to the SDW state, as determined by the peak of the derivative of the electrical resistance, together to the activation energy $E_a$, determined as in Fig 1 b, for different published *Ln-1111* compounds as a function of the basal lattice parameter, $La^{2, 3, 3233}$, $Gd^{34}$, $Ce^{3536}$, $Tb^{37}$, $Nd^{36}$, $Gd^{36}$, $Pr^{36}$ and $Sm$ $^{38}$,$^{39}$. The data shown with half-squares corresponds to structural and electrical resistance measurements on *Sm-1111* under pressure[40]. (b) Same as (a) for the *Æ-122* family, $Ba^5$, $Sr^{31}$, $Ca^{41}$. (c) Plot of the activation enegy as a function of the SDW transition temperature. Only *Sm-1111* [40] and *La-1111* [42] have been measured under pressure where each point corresponds to one pressure. (d) Same as (c) for the *Æ-122* compounds, $Ba^{43}$, $Sr^{44}$, $Ca^{41}$. The dependences are surprisingly linear and parallel within each family of compounds.

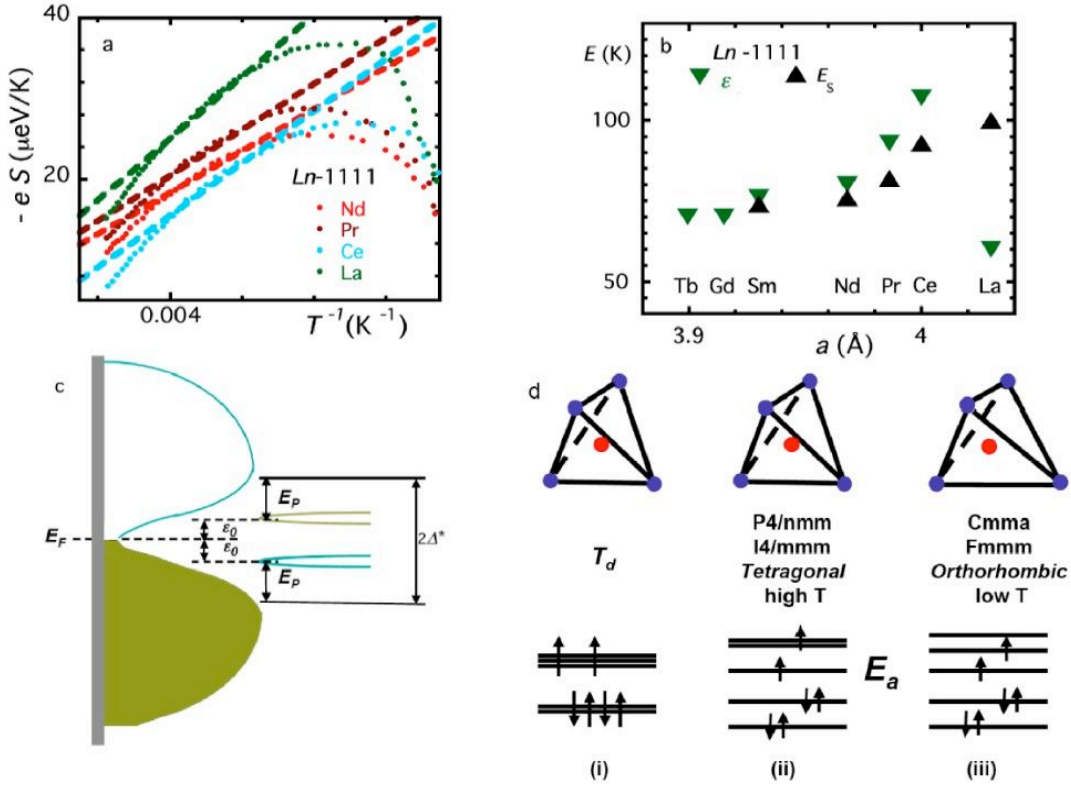

Figure 3 (a) Thermopower of *Ln-1111* compounds to determine the activation energy $E_S$. (b) Comparison of the intercept $\varepsilon$ from Fig. 2 c (the value for compounds where there are no pressure measurements, were obtained using the same slope as for *Sm*-1111 and *La*-1111) to $E_S$ as a function of the basal lattice parameter. For the $Sm^{38}$, $Nd^{32}$, $Pr^{32}$ and $Ce^{32}$ compounds the agreement is better than 20%. (c) Schematic band diagram for pnictides. The density of states derived from the conduction bands has a dip corresponding to the pseudogap $2\Delta^*$ that exists above $T^*$. Two narrow polaronic bands are shown, separated by an energy $2\varepsilon_0=2E_S$, corresponding to hole or electron polarons. $E_P$ is the formation energy of a polaron. (d) Levels at the *Fe* site for (i) a tetrahedral environment, (ii) the tetragonal high temperature structure and (iii) the low temperature orthorhombic structure[45]. The filling explains the JT effect, as the splitting of the degenerate level induced by the orthorhombic distortion causes an energy gain (ε>0), in the case of the *Ln*-1111 and *Ba*-122 compounds (for the other 122 materials the doublet and the singlet are probably inversed).

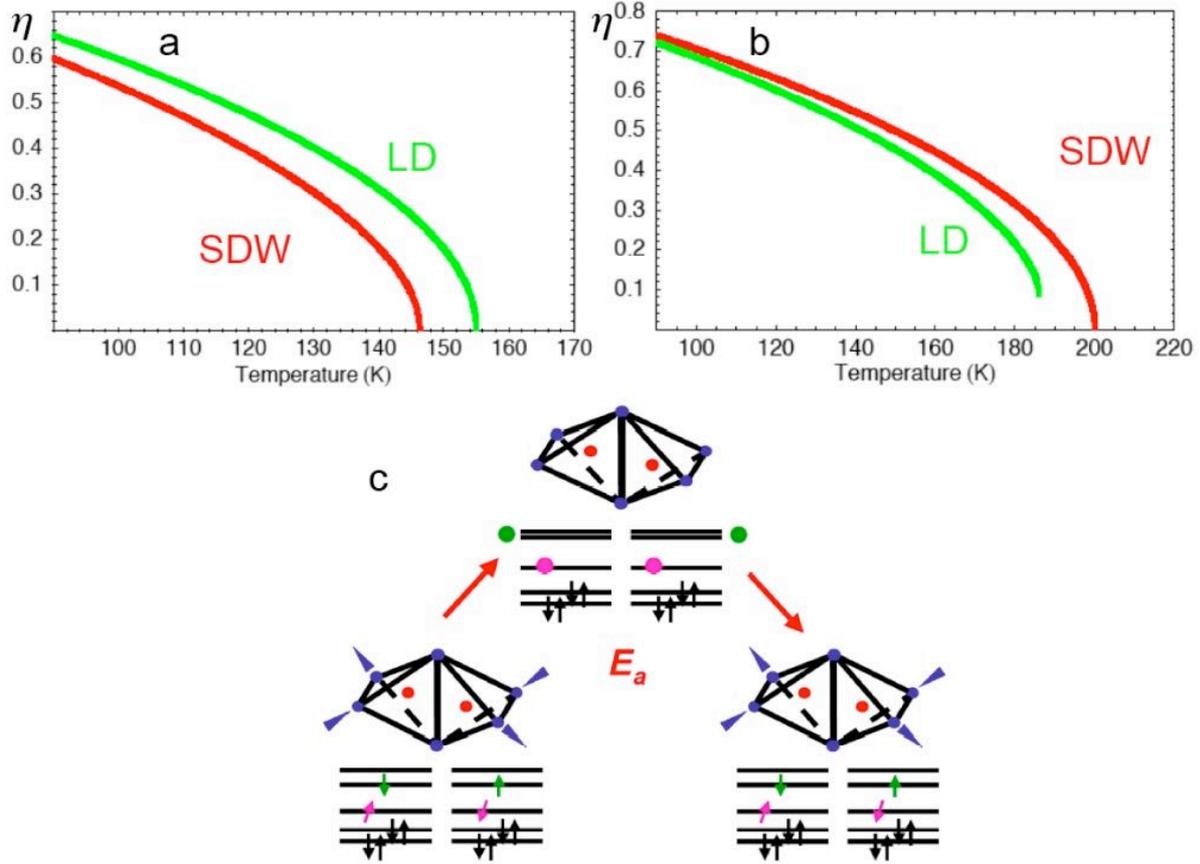

Figure 4 (a) Temperature dependence of the order parameter of the SDW (red) and the TOLD (green) according to the Landau free energy (1) for uncoupled transition temperatures $^{u}T_0(=155K) > T_L^*(=90K)$, both transitions are of second order and $T_0=155K$ and $T^*=145K$. (b) Same as (a) but for the case $^{u}T_L^*(=200K) > {^{u}T_0}(=0K)$. This is the strictest case, higher $^{u}T_0$ will give almost identical final transition temperatures, but for all $T^* > T_0$ cases the TOLD transition will be of first order. (c) Possible structure of the antiferromagnetic bipolaron and its conducting mechanism. The energy levels and spin disposition have been taken from Ref. 12. The bipolarons are in the deformed state and can be excited to the undeformed state through the excitation energy $E_a$, that frees the uppermost electrons (green circles) for a FC conduction though they are later pinned down to a new bipolaron state.